\documentclass[12pt]{article}
\usepackage{amssymb,amsmath,epsfig}
\allowdisplaybreaks

\begin{document}

\title{\bf Thermodynamics of Modified Cosmic Chaplygin Gas}
\author{M. Sharif \thanks {msharif.math@pu.edu.pk} and Sara Ashraf
\thanks{saraashraffarooq@gmail.com}\\
Department of Mathematics, University of the Punjab,\\
Quaid-e-Azam Campus, Lahore-54590, Pakistan.}

\date{}
\maketitle

\begin{abstract}
In this paper, we examine the thermodynamic features of an exotic
fluid known as modified cosmic Chaplygin gas in the context of
homogeneous isotropic universe model. For this purpose, the behavior
of physical parameters are discussed that help to analyze nature of
the universe. Using specific heat formalism, the validity of third
law of thermodynamics is checked. Furthermore, with the help of
thermodynamic entities, the thermal equation of state is also
discussed. The thermodynamic stability is explored by means of
adiabatic, specific heat and isothermal conditions from classical
thermodynamics. It is concluded that the considered fluid
configuration is thermodynamically stable and expands adiabatically
for an appropriate choice of parameters.
\end{abstract}

\section{Introduction}

The discovery of accelerated expansion of the universe has
unambiguously been proved by a diverse set of high-precision
observational data accumulated from various astronomical sources
\cite{1}. Dark energy (DE) is considered as the root cause behind
this tremendous change in cosmic history. It possesses negatively
large pressure which violates the strong energy condition
($p+3\rho<0$, where $p$ and $\rho$ are pressure and energy density,
respectively) but its complete characteristics are still not known.
Planck's observational data reveals that about $68.3$\% of our
universe is filled with this mysterious form of energy while other
cosmic budget includes $4.9$\% ordinary matter and $26.8$\% dark
matter (DM) \cite{2}. To explore the perplexing nature of DE, there
began a search for different candidates that can play their role as
an alternative for DE. The simplest candidate is the cosmological
constant while other favorable approaches include quintessence,
k-essence, Chaplygin gas (CG) etc known as DE matter models
\cite{3}.

Chaplygin gas is an intriguing model presented by a Russian
physicist Chaplygin as a convenient soluble model to study the
lifting force on the wing of an aeroplane in aerodynamics. It
efficiently describes the cosmic expansion and elegantly discusses
DM and DE in a unified form. The distinct feature of this model is
its positive and bounded squared speed of sound $(v_{s}^{2})$
leading to stable results as compared to other fluids with negative
pressure. Chaplygin gas acts as an alternative for dust dominated
era with small value of the scale factor and tends to cosmic
expansion for its large value while primordial universe cannot be
discussed in this scenario \cite{4}. Despite of the fact that it
does not meet the strong energy condition, it successfully shows
consistency with the observational results accumulated from various
cosmic probes such as Hubble space telescope, Wilkinson microwave
anisotropy probe, cosmic background explorer, etc.

To discuss the cosmic history as well as to get more accuracy with
observational data, several modifications of CG have been presented
which are obtained by introducing new parameters in its equation of
state (EoS). Bento et al. \cite{5} established the generalization of
this model named as generalized Chaplygin gas (GCG) which interprets
the same evolutionary phases of the universe as CG. Benaoum \cite{6}
introduced the modified Chaplygin gas (MCG) which illustrates the
radiation dominated era. Gonz$\acute{a}$lez-D$\acute{i}$az \cite{7}
proposed the generalization of cosmic CG models known as generalized
cosmic Chaplygin gas (GCCG) in such a way that it avoids big-rip
(singularity at a finite time) which was previously presented in the
DE models representing phantom era. This generalization provides
stable and physical behavior models even when the vacuum matter
configuration fulfills the phantom energy condition
($p+\rho<0,~\rho>0,~ \omega<-1$, where $\omega$ is the EoS
parameter). The other proposed CG models include variable Chaplygin
gas (VCG), variable modified Chaplygin gas (VMCG), new variable MCG,
extended CG and modified cosmic Chaplygin gas (MCCG) \cite{8}.

Chaplygin gas models have stimulated many researchers to investigate
their thermal stability. Santos et al. \cite{9} explored thermal
stability of GCG as well as MCG and deduced that these fluid models
verify the third law of thermodynamics along with the adiabatic
expansion. Myung \cite{10} proved the third law of thermodynamics
for CG model and illustrated that it can represent a unified picture
of DM and DE without any phase transition. Kahya and Pourhassan
\cite{11} analyzed extended CG model cosmologically as well as
thermodynamically and found stable results without any phase
transition against density perturbations. They also concluded that
all laws of thermodynamics are satisfied for this exotic fluid which
came out to be thermodynamically stable at all times. Panigrahi
\cite{12} studied thermodynamic behavior of VCG and observed that it
is thermodynamically stable throughout the evolution. He also
discussed thermal EoS which is an explicit function of temperature
only and checked the validity of the third law. Panigrahi and
Chatterjee \cite{13} found that current accelerated expansion of the
universe can be explained using VMCG model. Sharif and Sarwar
\cite{14} explored how GCCG can explain accelerated expansion of the
universe by interpreting different physical parameters and showed
that the fluid is adiabatically stable.

Here, we investigate thermodynamic stability of MCCG model in the
background of isotropic and homogeneous universe model. In section
\textbf{2}, we discuss the behavior of physical parameters such as
pressure, EoS as well as deceleration parameters and analyze the
stability using speed of sound. Section \textbf{3} deals with the
thermodynamic stability of MCCG. The results are summarized in the
last section.

\section{Physical Parameters for MCCG}

In this section, we discuss the behavior of MCCG in the background
of FRW universe model for different physical parameters and examine
its stability through squared speed of sound. The line element for
FRW universe model is given by
\begin{equation}\label{1}
ds^{2}=dt^{2}-a^{2}(t)(dr^{2}+r^{2}d\theta^{2}+r^{2}\sin^{2}\theta
d\phi^{2}),
\end{equation}
where $a(t)$ is the scale factor. The EoS for MCCG is defined as
\begin{equation}\label{2}
P = A\rho-\rho^{-\alpha}[(\rho^{\alpha+1}-C)^{-\gamma}+C],\quad
0<\alpha\leq1,~-b<\gamma<0,~b\neq1,
\end{equation} where
$C=\frac{Z}{\gamma+1}-1$ ($Z$ is an arbitrary constant) and $A$ is a
positive constant. This EoS reduces to GCCG as $A\rightarrow0$
\cite{14} while GCG is recovered in the limit $A\rightarrow0$ along
with $\gamma\rightarrow0$ \cite{5}. The energy density of the fluid
configuration is given by
\begin{equation}\label{3}
\rho=\frac{U}{V},
\end{equation}
where $U$ and $V$ represent the internal energy and volume,
respectively. Classical thermodynamics provides a useful
relationship among the quantities $U,~V$ and $P$ in the form
\begin{equation}\label{4}
\left(\frac{dU}{dV}\right) = -P.
\end{equation}
Using Eq.(\ref{2}) in the above expression, we obtain
\begin{equation}\label{5}
\frac{dU}{dV}+\frac{AU}{V}=\frac{V^{\alpha}}{U^{\alpha}}\left[C+\left(
\frac{U^{\alpha+1}}{V^{\alpha+1}}-C\right)^{-\gamma}\right],
\end{equation}
which is a nonlinear ordinary differential equation. Its solution is
given by
\begin{equation}\label{6}
U\thickapprox
V\left[\frac{(\frac{d}{V})^{(\alpha+1)(A+1)}(A+1)+C+(-C)^{-\gamma}}
{A+1+\gamma(-C)^{-\gamma-1}}\right]^{\frac{1}{\alpha+1}},
\end{equation}
where we have used the binomial expansion upto first order and $d$
is an integration constant which is either universal constant or a
function of entropy $(S)$. The above equation can also be written as
\begin{equation}\label{7}
U=V\left[\frac{\left(\frac{\varepsilon}{V}\right)^{M}+C+(-C)^{-\gamma}}
{A+1+\gamma(-C)^{-\gamma-1}}\right]^{\frac{1}{\alpha+1}},
\end{equation}
where $\varepsilon=d(A+1)^{\frac{1}{M}},~M=(\alpha+1)(A+1)$ and
$A+\gamma(-C)^{-\gamma-1}\neq-1$. It is clearly observed that
internal energy of the fluid can only be discussed when $\gamma$ is
a whole number between the above mentioned range for positive values
of $C$. Using Eqs.(\ref{3}) and (\ref{7}), the energy density of
MCCG becomes
\begin{equation}\label{8}
\rho=\left[\frac{\left(\frac{\varepsilon}{V}\right)^{M}+C+(-C)^{-\gamma}}
{A+1+\gamma(-C)^{-\gamma-1}}\right]^{\frac{1}{\alpha+1}}.
\end{equation}

In the following, we use this equation to discuss different physical
parameters.

\subsection{Pressure}
\begin{figure}
\centering \epsfig{file=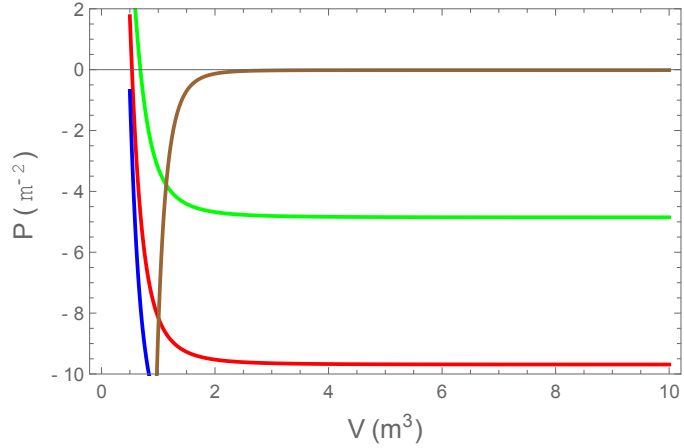,width=.65\linewidth}
\caption{Plots of $P$ versus $V$ for
$A=2,~\gamma=-2,~\alpha=0.1,~d=1$ with $Z=0.01$ (brown), $-5$
(green), $-7$ (red) and $-8$ (blue).}
\end{figure}

The pressure of MCCG in terms of $V$ can be obtained using
Eqs.(\ref{2}) and (\ref{8}) as
\begin{eqnarray}\nonumber
P&=&A\left(\frac{\left(\frac{\varepsilon}{V}\right)^{M}+C+
(-C)^{-\gamma}}{A+1+\gamma(-C)^{-\gamma-1}}\right)^{\frac{1}
{\alpha+1}}-\left(\frac{\left(\frac{\varepsilon}{V}\right)^{M}
+C+(-C)^{-\gamma}}{A+1+\gamma(-C)^{-\gamma-1}}\right)^{\frac{-\alpha}
{\alpha+1}}\\\label{9}&\times&\left[C+\left(\frac{\left(\frac
{\varepsilon}{V}\right)^M+C+(-C)^{-\gamma}}{A+1+\gamma(-C)^{-\gamma-1}}
-C\right)^{-\gamma}\right].
\end{eqnarray}
The graphical analysis of this equation is shown in Figure
\textbf{1} for different values of $Z$ with $A=2$. The positive and
negative behavior of pressure correspond to decelerated and
accelerated eras of the universe, respectively. For $Z=0.01$, it is
observed that the accelerating universe at small volume tends to
dust dominated universe at large volume. We note that the
decelerating universe tends to accelerate as volume increases for
$Z=-5,-7$ whereas accelerated phase is obtained only for $Z=-8$. The
same behavior of pressure is observed for different values of the
parameter $A$.

\subsection{EoS Parameter}

Here we discuss the effective EoS parameter of MCCG. Using
Eqs.(\ref{8}) and (\ref{9}), we have
\begin{equation}\label{10}
\omega=\frac{P}{\rho}=A-\frac{C+\left(\frac{\left(\frac{\varepsilon}{V}
\right)^{M}+C+(-C)^{-\gamma}}{A+1+\gamma(-C)^{-\gamma-1}}-C
\right)^{-\gamma}}{\left(\frac{\left(\frac{\varepsilon}{V}\right)^{M}
+C+(-C)^{-\gamma}}{A+1+\gamma(-C)^{-\gamma-1}}\right)}.
\end{equation}
We study the following two extremal cases for volume to analyze the
behavior of above equation.
\begin{itemize}
\item For small volume \emph{\texttt{$V\ll\varepsilon$}}, the
above equation reduces to
\begin{equation}\nonumber
P\approx A\rho,
\end{equation}
which is a barotropic EoS. In this case, $\omega$ will depend
entirely on the value of $A$.
\item For large volume \emph{\texttt{$V\gg\varepsilon$}}, Eq.(\ref{10})
takes the form
\begin{equation}\nonumber
\omega\approx\frac{P}{\rho}=A-\frac{C+\left(\frac{C+(-C)^{-\gamma}}
{A+1+\gamma(-C)^{-\gamma-1}}-C\right)^{-\gamma}}{\left(\frac{C+(-C)^{-\gamma}}
{A+1+\gamma(-C)^{-\gamma-1}}\right)}.
\end{equation}
\end{itemize}
For $\omega=0$, let volume be denoted by $V_{c}$, which is given by
\begin{equation}\nonumber
V_{c}=\varepsilon\left[\frac{A+\gamma(-C)^{-\gamma-1}}{C+(-C)^{-\gamma}}
\right]^{\frac{1}{M}}.
\end{equation}
The EoS parameter discusses both accelerated and decelerated phases
of the universe as well as successfully describes the phase
transitions (dubbed as flip) at a critical value $V_{c}$ between
these cosmic phases. The proper flip occurs when
$\left(\frac{\varepsilon}{V}\right)^{M}<1$ while the inequality
$C+(-C)^{-\gamma}\neq0$ leads to real flip.

Figure \textbf{2} shows the behavior of $\omega$ for different
values of $Z$. It is found that the value of $V_{c}$ decreases as
$Z$ becomes negatively large. The negative values of $Z$ show the
decelerated cosmic phase at small volume undergoing acceleration at
large volume while $Z=1$ demonstrates only the decelerating phase.
We observe that at large volume, the considered values of
$Z=1,~0,~-1.15$ and $-2$ correspond to stiff matter, dust dominated,
$\Lambda$CDM and phantom eras, respectively. Thus, the EoS parameter
can interpret different evolutionary phases of the universe.
\begin{figure}
\centering \epsfig{file=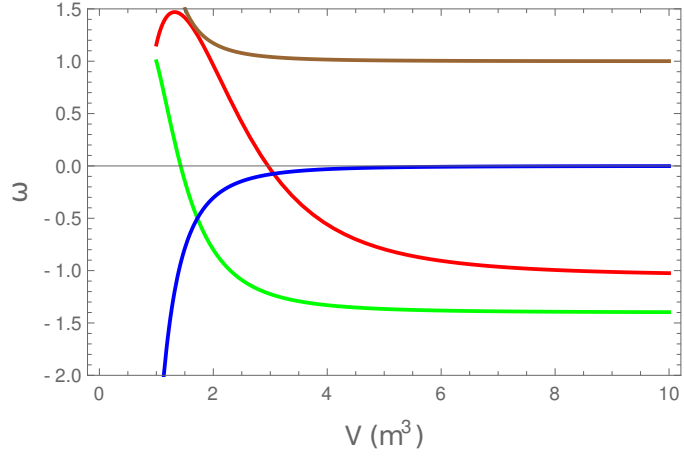,width=.65\linewidth} \caption{Plots
of $\omega$ versus $V$ for $\gamma=-2,~A=2,~\alpha=0.1,~d=1$ with
$Z=-2$ (green), $-1.15$ (red), $0$ (blue) and $1$ (brown).}
\end{figure}

\subsection{Deceleration Parameter}
\begin{figure} \centering
\epsfig{file=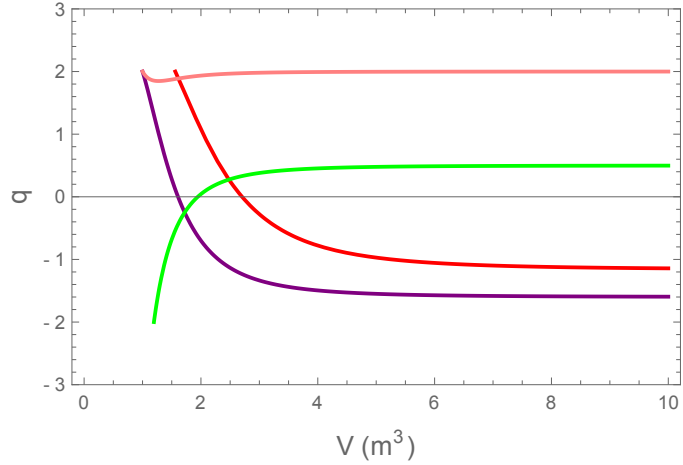,width=.65\linewidth} \caption{Plots of $q$ versus
$V$ for $\gamma=-2,~A=2,~\alpha=0.1,~d=1$ with $Z=-2$ (purple),
$-1.3$ (red), $0$ (green) and $2$ (pink).}
\end{figure}

The deceleration parameter is given by
\begin{equation}\label{11}
q=\frac{1}{2}+\frac{3P}{2\rho}.
\end{equation}
Using Eq.(\ref{10}), this parameter for MCCG takes the form
\begin{equation}\label{12}
q=\frac{1}{2}+\frac{3}{2}\left[A-\frac{C+\left(\frac{\left(
\frac{\varepsilon}{V}\right)^{M}+C+(-C)^{-\gamma}}{A+1+\gamma
(-C)^{-\gamma-1}}-C\right)^{-\gamma}}{\left(\frac{\left(\frac{
\varepsilon}{V}\right)^{M}+C+(-C)^{-\gamma}}{A+1+\gamma(-C)^{-\gamma
-1}}\right)}\right].
\end{equation}
For small volume, it reduces to
\begin{equation}\nonumber
q\approx\frac{1}{2}+\frac{3A}{2},
\end{equation}
which implies that the universe undergoes deceleration at its early
stage since $A>0$ while for large volume, Eq.(\ref{12}) becomes
\begin{equation}\nonumber
q\approx\frac{1}{2}+\frac{3}{2}\left[A-\frac{C+\left(\frac{C+(-C)^{-\gamma}}
{A+1+\gamma(-C)^{-\gamma-1}}-C\right)^{-\gamma}}{\left(\frac{C
+(-C)^{-\gamma}}{A+1+\gamma(-C)^{-\gamma-1}}\right)}\right].
\end{equation}
In this case, the flip occurs when deceleration parameter vanishes
and the corresponding flip volume $(V_{f})$ is given by
\begin{equation}\nonumber
V_{f}=\varepsilon\left[\frac{3A+1+3\gamma(-C)^{-\gamma-1}}{C
+(-C)^{-\gamma}}\right]^{\frac{1}{M}},
\end{equation}
provided that $C+(-C)^{-\gamma}\neq0$ and the inequality
$C+(-C)^{-\gamma}<3(A+\gamma(-C)^{-\gamma-1})+1$ leads to proper
flip. Figure \textbf{3} shows the evolution of deceleration
parameter against volume for different values of $Z$. At small
volume, the universe undergoes deceleration while accelerating
behavior is observed at large volume for considered negative values
of $Z$. The flip occurs at $V\approx1.6$ and $3$ for $Z=-2$ and
$-1.3$, respectively. For $Z=0$, the deceleration parameter switches
from acceleration to deceleration at $V_{f}\approx2.2$ while no flip
is observed for $Z=2$.

\subsection{Speed of Sound}

Here we analyze the stability of MCCG using speed of sound as
\begin{eqnarray}\nonumber
v_{s}^2=\left(\frac{\partial P}{\partial\rho}\right)_{S}
&=&A+\frac{\gamma(\alpha+1)}{\left(\frac{\left(\frac{\varepsilon}
{V}\right)^{M}+C+(-C)^{-\gamma-1}}{A+1+\gamma(-C)^{-\gamma-1}}
-C\right)^{\gamma+1}}+\frac{\alpha}{\left(\frac{\left(\frac{\varepsilon}
{V}\right)^{M}+C+(-C)^{-\gamma-1}}{A+1+\gamma(-C)^{-\gamma-1}}\right)}
\\\label{13}&\times&\left[C+\left(\frac{\left(\frac{\varepsilon}{V}
\right)^{M}+C+(-C)^{-\gamma-1}}{A+1+\gamma(-C)^{-\gamma-1}}-C
\right)^{-\gamma}\right],
\end{eqnarray}
whose feasible range is $0<v_{s}^{2}<1$. This equation reduces to
$v_{s}^2=A$ at early universe while for $V\gg\varepsilon$, we have
\begin{eqnarray}\nonumber
v_{s}^2&=&A+\frac{\gamma(\alpha+1)}{\left(\frac{C+(-C)^{-\gamma}}
{A+1+\gamma(-C)^{-\gamma-1}}-C\right)^{\gamma+1}}+\frac{\alpha}{\left(
\frac{C+(-C)^{-\gamma}}{A+1+\gamma(-C)^{-\gamma-1}}\right)}\\\nonumber
&\times&\left[C+\left(\frac{C+(-C)^{-\gamma}}{A+1+\gamma(-C)^{-\gamma}}
-C\right)^{-\gamma}\right].
\end{eqnarray}
Figure \textbf{4} shows the behavior of squared speed of sound
against the positive parameter $A$ for different values of $Z$. It
is observed that the viable ranges for $A$ are $1.8<A<2.8$,
$3.2<A<3.7$ and $5<A<5.6$ corresponding to $Z=-0.01$, $4$ and $7$,
respectively. Thus, the stable results are found for the considered
values of $Z$ in a particular range of $A$.
\begin{figure}
\centering \epsfig{file=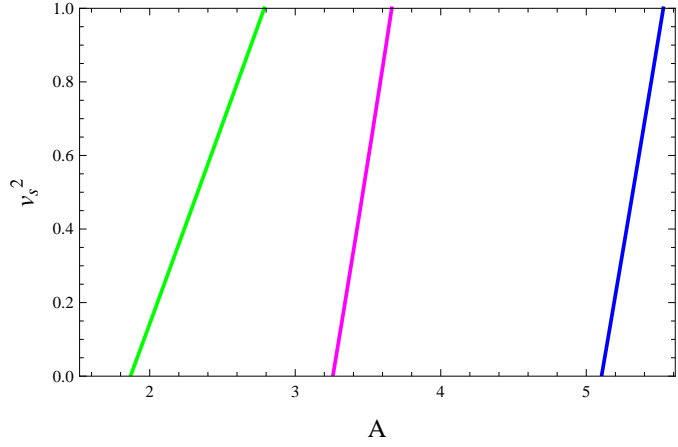,width=.65\linewidth} \caption{Plots
of $v_{s}^{2}$ versus $A$ for $\gamma=-2,~\alpha=0.1,~d=1$ with
$Z=-0.01$ (green), $4$ (magenta) and $7$ (blue).}
\end{figure}

\section{Thermodynamic Stability}

In this section, we discuss thermodynamic stability of MCCG during
its evolution. The stability conditions are given by \cite{15}
\begin{itemize}
\item The pressure reduces for both adiabatic as well as isothermal expansions as
\begin{equation}\label{14}
\left(\frac{\partial P}{\partial V}\right)_{S}<0,\quad
\left(\frac{\partial P}{\partial V}\right)_{T}<0,
\end{equation}
where $T$ represents temperature.
\item Specific heat at constant volume $(c_{V})$ is positive.
\end{itemize}
Differentiation of Eq.(\ref{9}) with respect to volume yields
\begin{eqnarray}\nonumber
\left(\frac{\partial P}{\partial V}\right)_{S}&=&
\frac{(A+1)\varepsilon^{M}}{V^{M+1}A+1+\gamma(-C)^{-\gamma-1}}
\left(\frac{\varepsilon^{M}V^{-M}+C+(-C)^{-\gamma}}{A+1+\gamma
(-C)^{-\gamma-1}}\right)^{\frac{-\alpha}{\alpha+1}}\\\nonumber&
\times &\left[-A(\alpha+1)+\alpha P
\left(\frac{\varepsilon^{M}V^{-M}+C+(-C)^{-\gamma}}{A+1+\gamma
(-C)^{-\gamma-1}}\right)^{\frac{-1}{\alpha+1}}\right.\\\label{16}
&-&\left.\gamma(\alpha+1)\left(\frac{\varepsilon^{M}V^{-M}+C+
(-C)^{-\gamma}}{A+1+\gamma(-C)^{-\gamma-1}}-C\right)^{-\gamma-1}\right].
\end{eqnarray}
When volume is very small, the above equation reduces to zero while
for large volume, we have the following expression
\begin{eqnarray}\nonumber
\left(\frac{\partial P}{\partial
V}\right)_{S}&=&\frac{(A+1)\varepsilon^{M}}{V^{M+1}(A+1+\gamma
(-C)^{-\gamma-1})}\left(\frac{C+(-C)^{-\gamma}}{A+1+\gamma
(-C)^{-\gamma-1}}\right)^{\frac{-\alpha}{\alpha+1}}\\\nonumber
&\times&\left[\alpha P\left(\frac{C+(-C)^{-\gamma}}
{A+1+\gamma(-C)^{-\gamma-1}}\right)^{\frac{-1}{\alpha+1}}-A(\alpha+1)
-\gamma(\alpha+1)\right.\\\label{18}&\times&\left.\left(\frac{C+
(-C)^{-\gamma}}{A+1+\gamma(-C)^{-\gamma-1}}-C\right)^{-\gamma-1}\right].
\end{eqnarray}
Figure \textbf{5} shows that the adiabatic condition is fulfilled
for all the considered values of $Z$.
\begin{figure}
\centering \epsfig{file=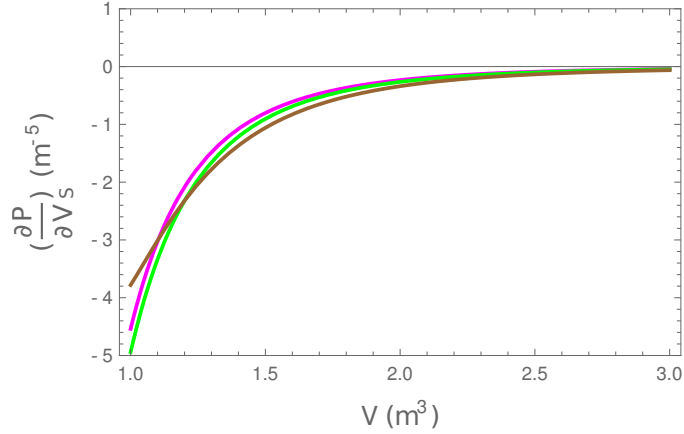,width=.65\linewidth}
\caption{Plots of $\left(\frac{\partial P}{\partial V}\right)_{S}$
versus $V$ for $\gamma=-2,~A=2,~\alpha=0.1,~d=1$ with $Z=-15$
(magenta), $-7$ (green) and $-2$ (brown).}
\end{figure}

To investigate the positivity of specific heat at constant volume,
we consider specific heat in terms of temperature and entropy as
\begin{equation}\label{19}
c_{v}=T\left(\frac{\partial S}{\partial T}\right)_{V},
\end{equation}
where the temperature of MCCG is obtained from the following
relation
\begin{equation}\label{20}
T=\frac{\partial U}{\partial S}=\left(\frac{\partial U}{\partial
d}\right)\left(\frac{\partial d}{\partial S}\right).
\end{equation}
Differentiating Eq.(\ref{6}) with respect to $d$, we have
\begin{equation}\label{21}
\frac{\partial U}{\partial d}=\frac{d^{M-1}(A+1)^{2}U}
{d^{M}(A+1)+CV^{M}+(-C)^{-\gamma}V^{M}}.
\end{equation}
Substituting this relation in Eq.(\ref{20}), the expression of $T$
becomes
\begin{equation}\label{22}
T=\frac{d^{M-1}(A+1)^{2}U}{d^{M}(A+1)+CV^{M}+(-C)^{-\gamma}V^{M}}
\left(\frac{\partial d}{\partial S}\right).
\end{equation}
When $d$ is a universal constant ($\frac{\partial d}{\partial
S}=0$), the temperature vanishes while it varies for CG expansion,
so we consider $\frac{\partial d}{\partial S}\neq0$. Here, we assume
the case $\frac{\partial d}{\partial S}>0$ to have a positive
temperature which is cooled down through adiabatic expansion. Using
the concept of dimensional analysis, Eq.(\ref{6}) gives
\begin{equation}\nonumber
[d]^{A+1}=[U][V]^{A}.
\end{equation}
Using the relation [$U$]=[$T$][$S$], the above equation becomes
\begin{equation}\label{23}
[d]=[T]^{\frac{1}{A+1}}[S]^{\frac{1}{A+1}}[V]^{\frac{A}{A+1}}.
\end{equation}
Taking $d=d(S)$, it follows that
\begin{equation}\label{24}
d=\left(\tau \nu^{A}S\right)^{\frac{1}{A+1}},
\end{equation}
where $\tau$ and $\nu$ are constants having the dimensions of
temperature and volume, respectively. Differentiating Eq.(\ref{24})
with respect to $S$, we obtain
\begin{equation}\label{25}
\frac{\partial d}{\partial S} = \frac{1}{A+1}\left(\frac{\tau
\nu^{A}}{S^{A}}\right)^{\frac{1}{A+1}}.
\end{equation}
Substituting this value in Eq.(\ref{22}), the temperature of MCCG
takes the form
\begin{equation}\nonumber
T=\frac{d^{M-1}(A+1)UB^{\frac{1}{A+1}}S^{\frac{-A}{A+1}}}
{d^{M}(A+1)+CV^{M}+(-C)^{-\gamma}V^{M}},
\end{equation}
where $B=\tau\nu^{A}$. Using Eqs.(\ref{6}) and (\ref{24}) in the
above equation, we have
\begin{equation}\label{26}
T=\frac{S^{\alpha}B^{\alpha+1}(A+1)
[(BS)^{\alpha+1}(A+1)+CV^{M}+(-C)^{-\gamma}V^{M}]^{\frac{-\alpha}{\alpha+1}}}
{V^{A}(A+1+\gamma(-C)^{-\gamma-1})^{\frac{1}{\alpha+1}}}.
\end{equation}

For a positive definite entropy, we assume $0<T<\tau$ and $0<V<\nu$.
It is worth mentioning here that when $T=0$, the entropy vanishes
which indicates that the considered fluid obeys third law of
thermodynamics (if the temperature of a physical system approaches
to zero then entropy becomes zero). Differentiating Eq.(\ref{26})
with respect to $S$, we obtain
\begin{eqnarray}\label{27}
\frac{\partial T}{\partial S}&=&\alpha ZS^{\alpha}
\left[S^{-1}\left\{(A+1)(BS)^{\alpha+1}+CV^{M}+(-C)^{-\gamma}V^{M}
\right\}^{\frac{-\alpha}{\alpha+1}}\right.\\\nonumber&-&\left.B^{\alpha
+1}S^{\alpha}(A+1)\left\{(A+1)(BS)^{\alpha+1}+CV^{M}+(-C)^{-\gamma}V^{M}
\right\}^{\frac{-2\alpha-1}{\alpha+1}}\right],
\end{eqnarray}
where
$Z=\frac{B^{\alpha+1}(A+1)}{V^{A}(A+1+\gamma(-C)^{-\gamma-1})^{\frac{1}
{\alpha+1}}}$. Inserting Eqs.(\ref{26}) and (\ref{27}) in
(\ref{19}), it follows that
\begin{equation}\label{28}
c_{V}=\frac{S}{\alpha}\left[1-\frac{(BS)^{\alpha+1}(A+1)}
{[(BS)^{\alpha+1}(A+1)+CV^{M}+(-C)^{-\gamma}V^{M}]}\right]^{-1}.
\end{equation}
For $c_{V}$ to be real,
$(BS)^{\alpha+1}(A+1)\neq(BS)^{\alpha+1}(A+1)+CV^{M}+(-C)^{-\gamma}V^{M}$
and to be positive, the inequality
$(BS)^{\alpha+1}(A+1)<(BS)^{\alpha+1}(A+1)+CV^{M}+(-C)^{-\gamma}V^{M}$
must hold. The graphical analysis of Eq.(\ref{28}) is shown in
Figure \textbf{6} for different values of $Z$ with $A=2$. We observe
that the positivity of specific heat is obtained for $Z=6$ in the
range $V>1.5$ while MCCG is thermally stable for both values of
$Z=0$ and $-0.5$ throughout the evolution. It is also noted that
when temperature is zero, thermal capacity vanishes which also
assures the validity of third law of thermodynamics.
\begin{figure}
\centering \epsfig{file=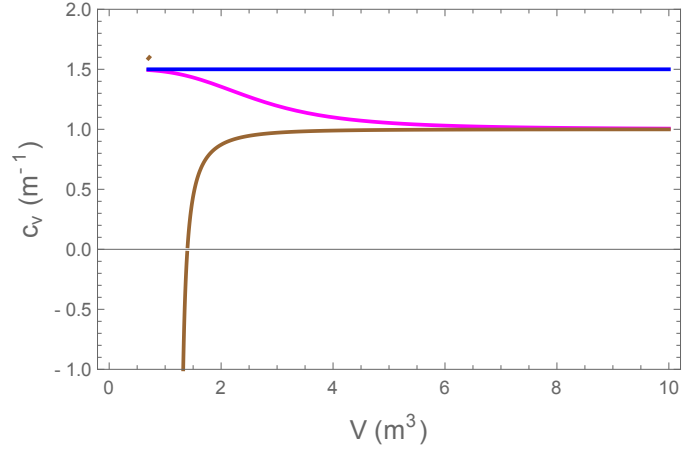,width=.65\linewidth} \caption{Plots
of $c_{V}$ versus $V$ for
$\gamma=-2,~A=2,~\alpha=0.1,~\tau=2.73,~S=1,~\nu=1,~d=1$ with
$Z=-0.5$ (magenta), $0$ (blue) and $6$ (brown).}
\end{figure}

Finally, we analyze the behavior of considered model through
isothermal condition. For this purpose, we assume $P=P(V,T)$ and by
solving Eqs.(\ref{9}) and (\ref{24}), we have
\begin{eqnarray}\nonumber
P&=&A\left(\frac{(BS)^{\alpha+1}(A+1)V^{-M}+C+(-C)^{-\gamma}}
{A+1+\gamma(-C)^{-\gamma-1}}\right)^{\frac{1}{\alpha+1}}\\\nonumber
&-&\left(\frac{(BS)^{\alpha+1}(A+1)V^{-M}+C+(-C)^{-\gamma}}
{A+1+\gamma(-C)^{-\gamma-1}}\right)^{\frac{-\alpha}{\alpha+1}}
\\\label{30}&\times&\left[C+\left(\frac{(BS)^{\alpha+1}(A+1)
V^{-M}+C+(-C)^{-\gamma}}{A+1+\gamma(-C)^{-\gamma-1}}
-C\right)^{-\gamma}\right].
\end{eqnarray}
The corresponding EoS parameter takes the form
\begin{equation}\label{31}
\omega=\frac{P}{\rho}
=A-\frac{C+\left(\frac{(BS)^{\alpha+1}(A+1)V^{-M}+C+(-C)^{-\gamma}}
{A+1+\gamma(-C)^{-\gamma-1}}-C\right)^{-\gamma}}{\left(\frac{(BS)
^{\alpha+1}(A+1)V^{-M}+C+(-C)^{-\gamma}}{A+1+\gamma(-C)^{-\gamma-1}}
\right)}.
\end{equation}
To check the isothermal condition, we should have $\rho=\rho(T)$ and
$P=P(T)$. In our case, it is difficult to have a thermal EoS for
MCCG as a function of temperature only since Eq.(\ref{26}) is a
complicated equation such that the explicit expression for $S$ in
terms of $T$ cannot be extracted. For this reason, we are unable to
analyze the isothermal condition in this scenario.

\section{Conclusions}

In this paper, we have analyzed thermodynamic stability of MCCG
within the framework of FRW universe model. We have examined the
expanding evolution of the universe through different physical
parameters like pressure, effective EoS and deceleration parameters
as well as speed of sound. The results of these parameters can be
summarized as follows.
\begin{itemize}
\item The consistent behavior of pressure with the evolutionary picture
of the universe is obtained for the considered values of $Z$ whereas
inconsistent evolution is observed for its positive values (Figure
\textbf{1}).
\item The EoS parameter for MCCG depicts that decelerated and
accelerated phases of our universe can be discussed for different
values of parameter $Z$ (Figure \textbf{2}). We have also calculated
the critical value at $\omega=0$ and found that its value increases
as $Z$ increases from its negative values to zero.
\item The evolution of deceleration parameter against volume gives
the decelerated universe when $V<V_{f}$ for negative values of $Z$
while accelerating behavior is observed when $V>V_{f}$. For positive
values of $Z$, we have found only deceleration while at $Z=0$,
acceleration occurs before the flip (Figure \textbf{3}).
\item We have analyzed the stability of MCCG through speed of sound and
obtained stable regions at large volume for the considered values of
$Z$ (Figure \textbf{4}). For VMCG model, the squared speed of sound
could be positive or negative \cite{13} whereas for GCCG model, the
stable regions do not exist in late universe \cite{14}.
\end{itemize}

Finally, we have investigated thermodynamic stability of considered
fluid configuration using adiabatic, isothermal and specific heat
conditions. We have found the validity of adiabatic as well as
positivity of specific heat for the considered values of $Z$
(Figures \textbf{5} and \textbf{6}). It is worth mentioning here
that third law of thermodynamics is obeyed for MCCG. We conclude
that MCCG expands adiabatically and the expansion is
thermodynamically stable for a suitable choice of the parameters.

\end{document}